\documentclass[10pt,twocolumn]{paper}
\usepackage{epsfig, graphicx}
\usepackage[english]{babel}
\usepackage[margin=1.5cm]{geometry}

\title{\center \rm \bf Nuclear Magnetic Resonance in Noncollinear Antiferromagnet Mn$_3$Al$_2$Ge$_3$O$_{12}$}
\author{\small \rm Aleksey M. Tikhonov$^a$\/\thanks{tikhonov@kapitza.ras.ru}, Nikolai G. Pavlov$^b$, Oleg G. Udalov$^c$\/\thanks{udalov@ipmras.ru}\\
\small $^a$ Kapitza Institute for Physical Problems, Russian Academy of Sciences, Moscow, Russia\\
\small $^b$Moscow Institute of Physics and Technology, Dolgoprudnyi, Moscow region, Russia\\
\small $^c$Institute for Physics of Microstructures, Russian Academy of Sciences, Nizhni Novgorod, Russia}

\begin{document}
\maketitle

\abstract{ \it The $^{55}$Mn nuclear magnetic resonance spectrum of noncollinear 12-sublattice antiferromagnet Mn$_3$Al$_2$Ge$_3$O$_{12}$ has been studied in the frequency range of $200 - 640$\,MHz in the external magnetic field ${{\bf H}\parallel[001]}$ at ${T=1.2}$\,K. Three absorption lines have been observed in fields less than the field of the reorientation transition $H_c$ at the polarization ${{\bf h}\parallel{\bf H}}$ of the rf field. Two lines have been observed at $H > H_c$ and ${{\bf h} \perp {\bf H}}$. The spectral parameters indicate that the magnetic structure of manganese garnet differs slightly from the exchange triangular 120-degree structure. The anisotropy of the spin reduction and (or) weak antiferromagnetism that are allowed by the crystal symmetry lead to the difference of $\approx 3\%$ in the magnetization of sublattices in the field $H < H_c$. When the spin plane rotates from the orientation perpendicular to the $C_3$ axis to the orientation perpendicular to the $C_4$ axis, all magnetic moments of the electronic subsystem decrease by $\approx 2\%$ from the average value in the zero field.}

\vspace{0.25in}

The triangular 12-sublattice antiferromagnetic ordering is implemented in manganese garnet
Mn$_3$Al$_2$Ge$_3$O$_{12}$ at $T < 6.8$\,K. In this work, the garnet magnetic structure we report is studied using the nuclear magnetic resonance (NMR) spectra of $^{55}$Mn. Information about the relativistic distortions of the exchange spin structure of Mn$_3$Al$_2$Ge$_3$O$_{12}$ owing to the spin reduction anisotropy and weak antiferromagnetism is obtained from these spectra.

According to the neutron diffraction studies, the magnetic moments of Mn$^{2+}$ (the ground state $^6S_{5/2}$) in the magnetic ordered state of Mn$_3$Al$_2$Ge$_3$O$_{12}$ (crystallographic symmetry group O$^{10}_h$) lie in the (111) plane and are directed along or contrary to the [-211], [1-21],
and [11-2] axes (see Fig. 1) [1, 2]. When the external magnetic field ${\bf H}$ is applied along the [001] direction, the spin plane is rotated until the critical external field ${H_c\approx 2.4}$\,T is achieved [3]. The spin plane is perpendicular to the external field if its magnitude is higher than $H_c$.

\begin{figure}
\hspace{0.3in}
\epsfig{file=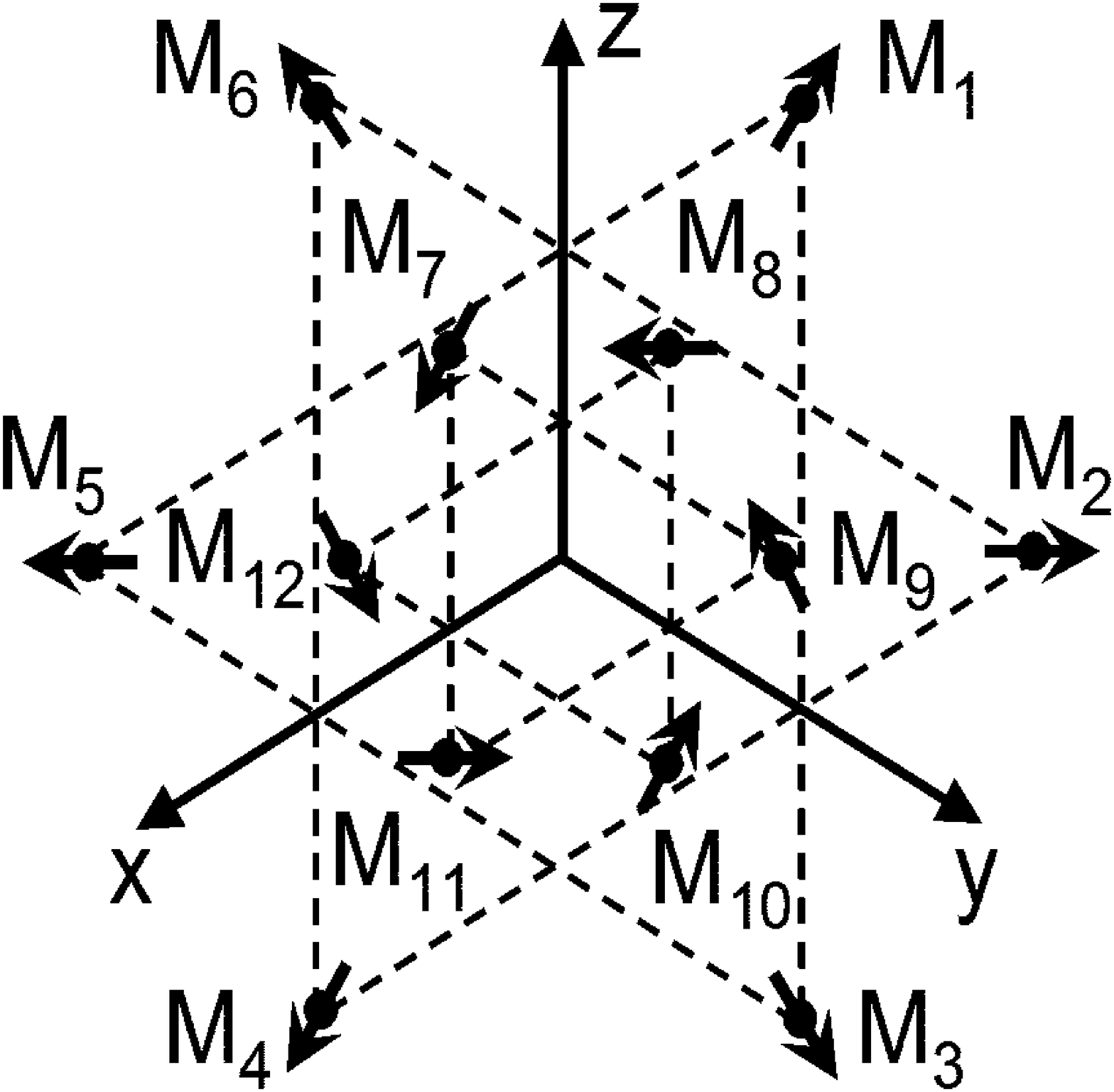, width=0.3\textwidth}

\small {\bf Figure 1}. Magnetic structure of noncollinear Mn$_3$Al$_2$Ge$_3$O$_{12}$ antiferromagnet in the exchange approximation.
\end{figure}

Three branches of the antiferromagnetic resonance were earlier observed in the study of Mn$_3$Al$_2$Ge$_3$O$_{12}$ at liquid helium temperatures in fields less than $H_c$ [4].
The frequency of one of them decreases strongly at ${{\bf H}\parallel[001]}$ when $H_c$ is approached. The exchange approximation was used to describe the results obtained in [4]. Since the experimental study of the
magnetic resonance was performed at frequencies higher than 20\,GHz, the results of this study were
interpreted disregarding the relativistic distortions of the exchange spin structure and the hyperfine interaction in the Mn$^{2+}$ ion.

The effect of the relativistic interactions in the Mn$_3$Al$_2$Ge$_3$O$_{12}$ crystal including the hyperfine interaction on the low-frequency part of the magnetic resonance spectrum was theoretically considered in [5]. To determine the theoretical constants and to test the theoretical results, we studied the $^{55}$Mn NMR spectrum (100\% isotopic composition) in Mn$_3$Al$_2$Ge$_3$O$_{12}$ crystals using a broadband continuous-wave NMR spectrometer, the functioning of which was described in [6].

In our experiments, we used a "split-ring" broad-band resonance system. 
It was placed in a vacuum chamber (the pressure of the heat-exchange $^4$He gas in
the chamber was less than 1 Torr) inserted in a liquid helium bath along with the garnet single-crystal 
sample. The temperature in the experiment $(T \approx 1.2 K)$ was controlled by the pressure of the saturated helium vapors in the bath. Absorption was recorded by magnetic field scanning at the fixed frequency $\nu$ of the rf field ${\bf h}$ (${\Delta\nu/\nu \sim 10^{-5}}$). 
We note that the feature of the observation of nuclear magnetic resonance in compounds with Mn$^{2+}$ ions is attributed to the high frequency of the resonance in zero field (${\nu_n {\sim 600}}$\,MHz).
This frequency is determined by the high value of the mean local field on the nucleus (${H_n\approx\nu_n/\gamma_n \sim 60}$\,T) caused by the hyperfine interaction between the
nuclear and ion spins (the gyromagnetic ratio for $^{55}$Mn ${\gamma_n \approx
10.57}$\,MHz/T). In addition, the dynamic shift of the NMR frequency is often observed in compounds
with magnetic Mn$^{2+}$ ions [7-9]. Therefore, the nuclear resonance can be observed in the frequency
range from 200 to 700\,MHz.

\begin{figure}
\hspace{0.3in}
\epsfig{file=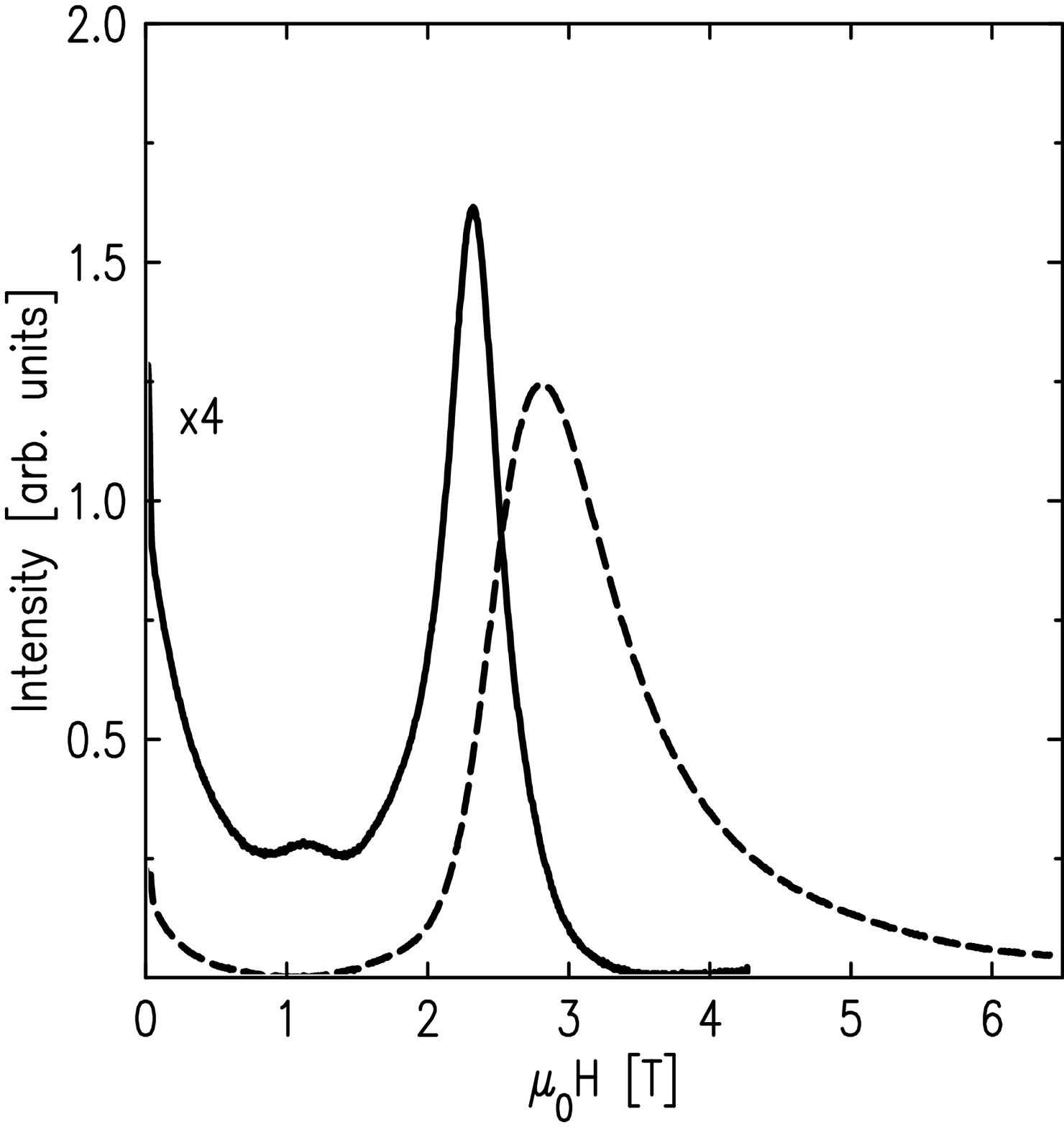, width=0.35\textwidth}

\small {\bf Figure 2}. Examples of magnetic field scans at a frequency of
${\sim 610}$\,MHz at (solid line) ${{\bf h}\parallel{\bf H}}$ and (dashed line)${{\bf h} \perp {\bf H}}$ (${{\bf H}\parallel[001]}$).

\end{figure}

Figure 2 shows the examples of field scans (${\bf H}\parallel[001]$) at the frequency ${\nu\sim 610}$\,MHz in the case of two polarizations ${{\bf h} \parallel{\bf H}}$ and ${{\bf h} \perp {\bf H}}$ (solid and dotted lines, respectively). Three absorption lines, which are excited by the rf field ${{\bf h}\parallel{\bf H}}$, were observed in the fields less than the field of the reorientation transition (${H_c\approx 2.4}$\,T). Two absorption lines which were excited at ${{\bf h}\perp{\bf H}}$ were observed when ${H>H_c}$. The absorption
intensity in this range strongly depends on the frequency. Its maximum is at ${\sim 608}$\,MHz. Figures 3 and 4 illustrate absorption maps in the linear regime for the frequency range 560 -- 640\,MHz when ${{\bf h}\parallel{\bf H}}$ and ${{\bf h} \perp {\bf H}}$, respectively. The nonlinear regime for the absorption in the garnet is discussed in [10].

Two absorption lines are observed at the external magnetic field directed along the [001] direction in the
frequency range of ${200 - 600}$\,MHz in the vicinity of $H_c$:
one at ${{\bf h}\parallel{\bf H}}$ and ${H<H_c}$ and the other at ${{\bf h} \perp {\bf H}}$ and ${H>H_c}$. The appearance of these branches in the wide frequency range is obviously due to the interaction
between the vibrations of the nuclear magnetization of $^{55}$Mn and the low-frequency branch of the antiferromagnetic resonance. Circles in Fig. 5 show the positions of the absorption maxima for the spectrum of the nucleus-like vibrations in Mn$_3$Al$_2$Ge$_3$O$_{12}$ at ${{\bf h}\parallel{\bf H}}$ and .

\begin{figure}
\hspace{0.3in}
\epsfig{file=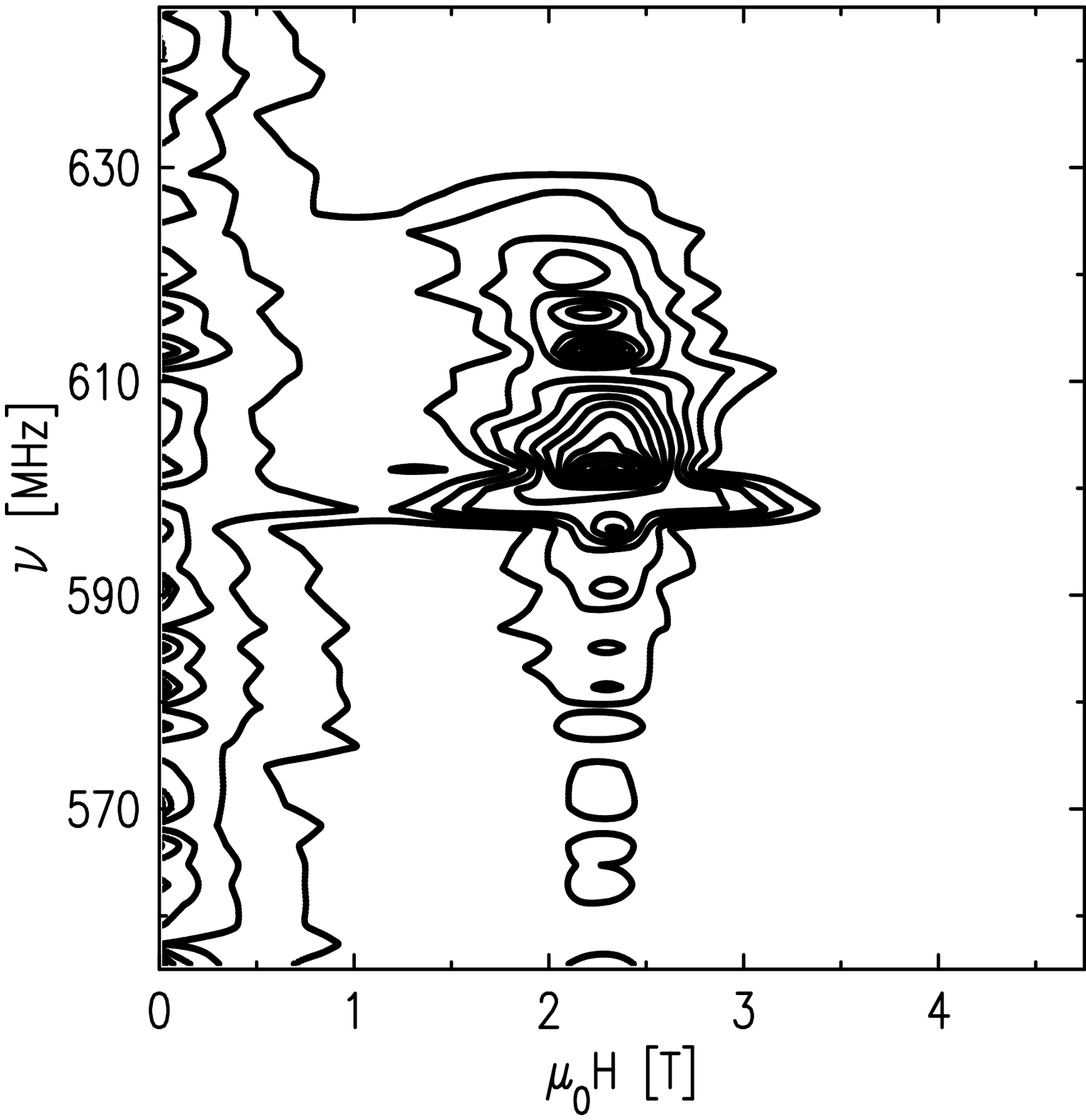, width=0.35\textwidth}

\small {\bf Figure 3}. Absorption map in the frequency range 560 -- 640\,MHz and ${\bf H} || [001]$
in antiferromagnet Mn$_3$Al$_2$Ge$_3$O$_{12}$ at ${\bf h} || {\bf H}$ ($T=1.2$\,K).

\end{figure}

To explain the experimental dependences of the resonance frequencies on the external magnetic field,
we used the theory developed in [5] with the application of the macroscopic approach [11]. The NMR frequencies are determined both by the energy of the hyperfine interaction between the magnetic moments
$m$ of $^{55}$Mn nuclei and the magnetic moments ${M= \gamma h \langle S \rangle}$ of electrons ${ -A({\bf m}, {\bf M})/h^2\gamma_n\gamma}$ (where ${\langle S\rangle}$ is the average electron spin in the ion,${h\approx 6.626\cdot 10^{-34}}$\,J$\cdot$s is the Planck constant, and ${\gamma\approx 28.02}$\,GHz/T is the electron gyromagnetic ratio) and by the interaction with the external magnetic field, ${-({\bf m},{\bf H})}$. The hyperfine interaction constant for the Mn$^{2+}$ ion in the oxygen environment is ${A\approx 1.6\cdot 10^{-25}}$\,J [12-14]. The frequency region in which the resonances of the nuclear
magnetic moments lie is determined by the mean hyperfine interaction field, ${H_n\approx A\langle S\rangle/h\gamma_n}$, since $H_n>>H$. It follows from our experiment that ${\nu_n \approx 615}$\,MHz,
i.e., ${\langle S\rangle \approx h\nu_n/A \approx 2.5}$.

The anisotropy in the Mn$_3$Al$_2$Ge$_3$O$_{12}$ crystal (when the external field is applied along the $C_{4}$ axis) may lead to two types of the distortions of the exchange spin garnet structure (see Fig. 1), which is described by two orthogonal unit vectors forming the spin plane [4]. First, there is the anisotropy of the spin reduction which was earlier observed in CsMnI$_{3}$ [15, 16]. In garnet, it is manifested in the relative change in the lengths of the orthogonal antiferromagnetic vectors of the order parameters ${\bf L}_1$ and ${\bf L}_2$ (${L^2_1\neq L^2_2}$). In the fields exceeding the field of the reorientation transition, this leads to canting of the magnetic moments of electrons with respect to the [–110] axis. Weak antiferromagnetism is the second type of distortions allowed by the symmetry of this crystal at this direction of the external field. The weak antiferromagnetism vector $\bf a$ is added to the magnetizations ${\bf M}_i$ of sublattices with ${i=1,\,2,\,3,\,10,\,11,\,12}$ (${{\bf M}_i\to{\bf M}_i+\bf a}$) and it is subtracted from the magnetizations with ${i=4,\,5,\,6,\,7,\,8,\,9}$ (${{\bf M}_i\to{\bf M}_i-\bf a}$) ${(a<<M)}$. The vector a is also directed along the [–110] axis (see Fig. 1). It is obvious that both phenomena lead to the deviation of the garnet magnetic structure from the triangular 120-degree structure.

\begin{figure}
\hspace{0.3in}
\epsfig{file=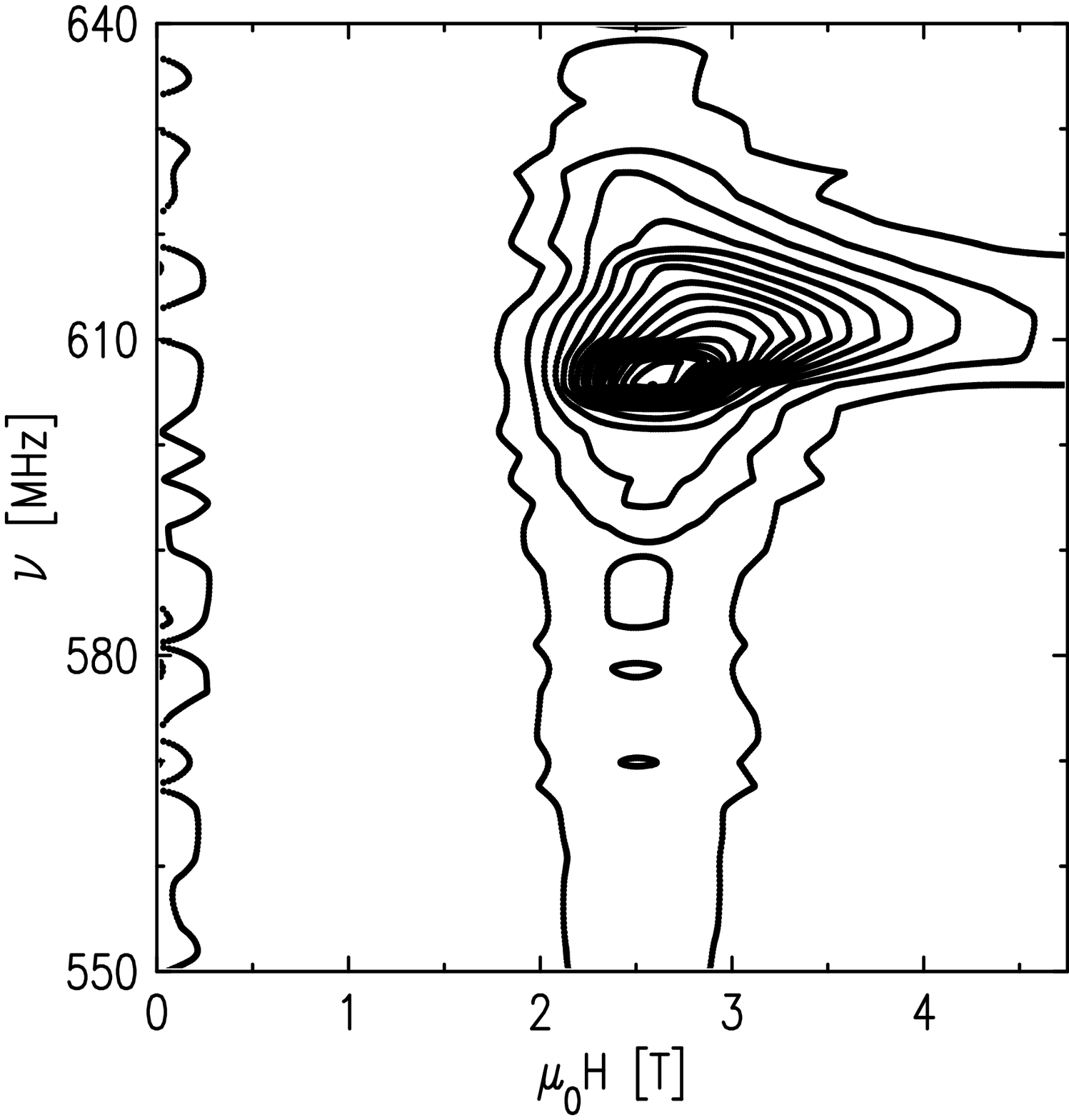, width=0.35\textwidth}

\small {\bf Figure 4}. Absorption map in the frequency range 560 -- 640\,MHz and ${\bf H} || [001]$
in antiferromagnet Mn$_3$Al$_2$Ge$_3$O$_{12}$ at ${\bf h} \perp  {\bf H}$ ($T=1.2$\,K).

\end{figure}

According to the theory, three NMR frequencies may be observed in zero field. The highest resonance
frequency is ninefold degenerate, the central frequency is not degenerate, and the lowest resonance
frequency is twofold degenerate. The distance between them is determined by the ratio ${\chi_n A^2M^2/\lambda_b}$. Here, $\lambda_{b}$ is the anisotropy energy and ${\chi_n = N_A\gamma_n^2h^2{\rm I(I+1)}/3kT}$ is the nuclear magnetic susceptibility, where ${N_A\approx 6.022\cdot 10^{23}}$\,mol$^{–1}$ is the Avogadro number, $I = 5/2$ is the spin of the $^{55}$Mn nucleus, and ${k\approx 1.381 \cdot10^{-23}}$\,J/K is the Boltzmann constant (${\chi_n \approx 5\cdot 10^{-7}}$\,sgs unit/mol at ${T=1.2}$\,K).

The other characteristic feature of the NMR spectrum is a "gap" in the field region of the reorientation
transition. This gap is due to the fact that one of the branches of the antiferromagnetic resonance spectrum
"descends" to the low-frequency region near the field $H_c$. This leads to the strong interaction between the electron and nuclear degrees of freedom and finally to the repulsion of one of the nuclear branches from the
electron one. The width (in the field) of the gap in the external field dependence of the resonance frequencies is proportional to the splitting of the spectral branches in zero field. By fitting the theoretical curves to the experimental ones in the gap region, the splitting of the NMR spectrum in zero field can be estimated as ${\sim 2}$\,MHz ($0.3\%$ of $\nu_n$).

The difference in the NMR frequencies in the external field is due both to the difference in the magnetic moments of electrons at different sites of the lattice and to the difference in their orientations with respect to the external field. Owing to the high symmetry of the Mn$_3$Al$_2$Ge$_3$O$_{12}$ crystal, the value and orientation of magnetic moments of electrons are determined by relatively few parameters. Some of these parameters are known from [4], where the vibration spectrum of the magnetic moments of electrons was
studied. In particular, the ratio of the longitudinal magnetic susceptibility of electrons (along the $[{\bf l}_1$\,${\bf l}_2]$ vector) to the transverse one is $\approx 1.28$ and the value ${\gamma(2\lambda_{b}/{\sqrt 3}\chi_{ \perp})^{1/2}=39}$\,GHz (here, ${\chi_{\perp} \approx 0.3}$\,cm$^3$/mol is the transverse magnetic susceptibility of the electronic subsystem [3] and ${\lambda_{b}\approx 4.4}$\,J/mol).

It was noted in [5] that the NMR spectrum at the orientation of the external field along the $C_4$ axis (the
[001] direction) generally contains nine branches, three of which are twofold degenerate. Two resonance
frequencies depending weakly on the external field are observed in the field region higher than the field of the reorientation transition. One of them corresponds to the branch rising from the low-frequency region. The second frequency corresponds to the other eight branches, which are not resolved in the experiment
owing to the weak splitting.

We note that the resonance in the field region ${H>H_c}$ lies ${5\div10}$\,MHz lower than the resonance at the zero field because a change in the electron magnetic moment caused by the relativistic distortions of the exchange structure is the same for all sites (see Eq. (6) in [5]). Lines in Fig. 5 show the fitting theoretical curves for the NMR spectrum in Mn$_3$Al$_2$Ge$_3$O$_{12}$ obtained when the weak antiferromagnetism value and spin reduction anisotropy were specified by two and one parameter, respectively.

Our data indicate that the distortions of the exchange triangular garnet structure connected with
the relativistic interactions in the crystal are small. It is possible to establish by fitting the theoretical curves to the experimental ones that the change in all magnetic moments of the electronic subsystem at the rotation of the spin plane from the orientation perpendicular to the $C_3$ axis to the orientation perpendicular to the $C_4$ axis is .${\approx 2\%}$ ${(\Delta M/M \approx -0.02)}$ and also to estimate the value of the exchange constant ($C_{1}\approx 250$\,J/mol) that determines the square of the order parameter (see Eqs. (4) and (6) in [5]). Unfortunately, it is not possible to reliably separate contributions to the splitting of the spectrum of weak antiferromagnetism and anisotropic spin reduction from the available experimental data in the field ${H>H_c}$.

\begin{figure}
\hspace{0.3in}
\epsfig{file=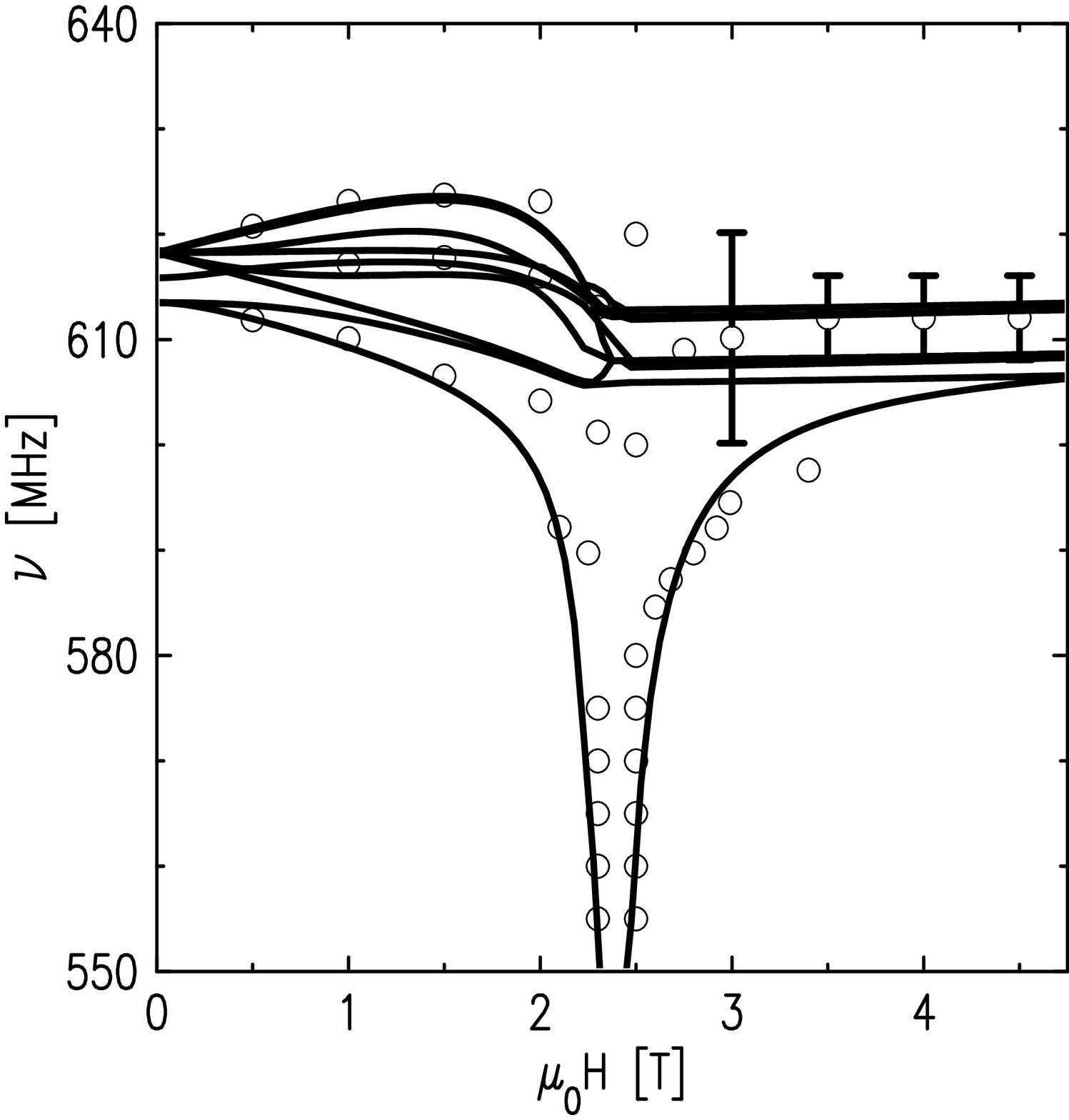, width=0.35\textwidth}

\small {\bf Figure 5}. Nuclear magnetic resonance spectrum in antiferromagnet Mn$_3$Al$_2$Ge$_3$O$_{12}$: circles are the positions of the absorption maxima in the experiment and lines are theoretical calculations [5].

\end{figure}

We are grateful to B.\,V.\,Mill for manganese garnet single crystals, to A.\,Yu.\,Semanin for the help in the experiments, and to V.\,I.\,Marchenko for helpful discussions.

\end{document}